\documentclass[conference]{IEEEtran}
\IEEEoverridecommandlockouts

\usepackage[utf8]{inputenc}

\usepackage[dvipsnames,svgnames,x11names]{xcolor}

\usepackage{xspace}

\usepackage{pifont}

\usepackage[normalem]{ulem}

\usepackage{amsmath}

\usepackage{amsfonts}

\usepackage{subcaption}

\usepackage{wrapfig}

\usepackage{adjustbox}

\usepackage{multirow}

\usepackage{makecell}

\usepackage{tabularx}

\usepackage{breakcites}

\usepackage[hyphens]{xurl}

\usepackage{comment}

\usepackage{lipsum}

\usepackage{graphicx}

\usepackage{booktabs}

\usepackage{enumitem}

\usepackage[normalem]{ulem}

\usepackage{tikz}
\usetikzlibrary{shapes.geometric}

\newcommand{\yes}{\textcolor{green!55!black}{\ding{51}}}
\newcommand{\no}{\textcolor{red!75!black}{\ding{55}}}
\newcommand{\pt}{\textcolor{orange!85!black}{$\circ$}}

\def\BibTeX{{\rm B\kern-.05em{\sc i\kern-.025em b}\kern-.08em
    T\kern-.1667em\lower.7ex\hbox{E}\kern-.125emX}}
\begin{document}

\pagenumbering{arabic}
\newcommand{\brand}{LLQM\xspace}

\title{Towards Continuous Profiling and Optimization of Quantum-Classical Pipelines
}

\author{\IEEEauthorblockN{Ayush Bansal$^{1}$, Owen Cochell$^{1}$, Santiago N\'{u}\~{n}ez-Corrales$^{1}$, Marcos Frenkel$^{1}$,\\
Seetharami Seelam$^{2}$, Apoorve Mohan$^{2\, *}$, Tianyin Xu$^{1}$}
\IEEEauthorblockA{\textit{$^{1}$University of Illinois Urbana-Champaign}\quad $^{2}$\textit{IBM Research}}
\thanks{$^{*}$Project Co-PI and co-advisor.}
}

\maketitle

\thispagestyle{plain}
\pagestyle{plain}

%
%

\newcommand{\para}[1]{\smallskip\noindent {\bf #1} }

\iftrue
        \makeatletter
        \def\UTFviii@defined#1{%
        \ifx#1\relax
                {\color{Magenta} $\boxtimes$}%
        \else
                \expandafter #1%
        \fi
        }
        \makeatother

        \newcommand{\ABadd}[1]{{\color{OliveGreen} \textit{#1}}} 
        \newcommand{\ABdel}[1]{{\color{Violet} \sout{#1}}} 
        \newcommand{\ABcmt}[1]{{\color{RoyalBlue} /* \textit{#1} */}} 

        \newcommand{\ab}[1]{{\ABadd{#1}}} 
        \newcommand{\abr}[2]{{\ABdel{#1} \ABadd{#2}}} 
        \newcommand{\abrc}[3]{{\ABdel{#1} \ABadd{#2} \ABcmt{#3}}} 

        \newcommand{\tianyin}[1]{{\color{Crimson} #1}} 
        \newcommand{\apoorve}[1]{{\color{Blue} #1}} 
        \newcommand{\owen}[1]{{\color{pink} #1}} 

        \let\oldcite\cite
        \renewcommand{\cite}[1]{\mbox{\oldcite{#1}}}

        \newcommand{\placeholder}[1]{{\color{Grey} \lipsum[1-#1]}}
    	\newcommand{\tocite}[1]{\textcolor{red}{**[cite: \@#1]**}}
    	\newcommand{\todo}[1]{\textcolor{DarkOrchid}{**[TODO: \@#1]**}}
\fi

\newcommand*\circled[1]{\tikz[baseline=(char.base)]{
            \node[shape=ellipse, draw, inner sep=2pt, scale=0.8, transform shape] (char) {#1};}}

\begin{abstract}

Quantum applications increasingly execute as multi-stage quantum-classical pipelines, 
    interleaving QPU computation with classical stages such as circuit generation, 
    transpilation, layout mapping, quantum error mitigation (QEM), and post-processing. 
These stages have diverse resource requirements and exhibit stochastic behavior under
drifting hardware noise, 
    yet existing workflow frameworks treat them as static, isolated components.
We present LLQM (Low-Level Quantum Machine), a profiling-driven meta-framework for
quantum-classical pipelines. 
LLQM decomposes pipelines into fine-grained tasks and continuously profiles their CPU/GPU, 
    memory, QPU, and queue dependencies alongside real-time hardware states. 
This unified runtime abstraction captures cross-stage resource dependencies 
    and reveals how classical and quantum decisions interact, 
    enabling characterization of their impact on fidelity and resource consumption.
We evaluate LLQM using QEM as a representative pipeline stage, on IBM 156-qubit Heron r2 processors with circuits 
    up to $\mathbf{100}$ qubits and $\mathbf{10^7}$ transpiled gates. 
Our results show that continuous profiling exposes runtime bottlenecks and enables hardware-, 
    fidelity-, and workload-aware optimizations.

\end{abstract}


\section{Introduction}
\label{sec:intro}


Quantum applications have evolved from small-scale experiments to utility-scale computations
    for quantum
    chemistry~\cite{ibm-riken-chem-sqd,ibm-lockheed-martin}, 
    drug discovery~\cite{ibm-cleveland-clinic-cross12000,ibm-cleveland-clinic-protein}, 
    and many-body dynamics~\cite{ibm-berkeley-physics,google-phases-of-matter,google-anyons}.
These applications span tens to hundreds of 
    physical qubits~\cite{quantum-supremacy,strong-quantum-advantage,ibm-berkeley-physics} 
    on quantum processors 
    such as the IBM Nighthawk (120 qubits)~\cite{ibm-nighthawk}, 
    IBM Heron (156 qubits)~\cite{ibm-heron}, and
    Google's Willow (105 qubits)~\cite{google-willow}. 
Current-generation Quantum Processing Units (QPUs) suffer from
    decoherence, gate imperfections, and readout errors that
    compound exponentially with circuit depth and width~\cite{qc-nisq,qc-large-circuits,ibm-qem-scalability}. 
Producing a usable result therefore
    takes more than executing a circuit---an application runs as a multi-stage quantum-classical
    pipeline, where quantum computation is interleaved with classical stages, 
    including circuit generation, transpilation, qubit mapping, error mitigation, 
    and post-processing. 
    
Different stages of quantum-classical pipelines exhibit distinct computational characteristics, 
    resource requirements, and optimization objectives, 
    which collectively determine the fidelity and the cost of computation.
Moreover,
stages are not fixed in their own behavior. As we demonstrate
in this work using Quantum Error Mitigation (QEM)~\cite{qem-google,em-short-depth-ibm,pec-nisq,prx-qem}, 
different techniques within the same pipeline stage exhibit drastically
different classical resource requirements, interact differently with the underlying quantum hardware,
and produce different execution fidelity. 
In this paper, we focus on QEM, which is
    resource-demanding and tightly coupled to the quantum hardware state.
Its optimization depends jointly on workload characteristics, classical resources, 
    and the evolving hardware state, rather than on any single factor in isolation.

Decomposing a workload into phases and profiling them individually 
    is a well-established practice in classical
    high-performance computing (HPC). 
There, a profile is taken against stationary hardware---the same code
on the same node next week exhibits the same bottleneck; hence, 
    one characterization serves many executions. 
However, these HPC practices do not directly apply to quantum workloads.
QPU error rates drift on timescales of minutes~\cite{tracking-drift}.
A noise model learned at the start of a sampling campaign can be invalid by its end; 
a qubit layout scored against a calibration snapshot can be stale before the job leaves the queue.
The cost to recover fidelity is a function of the noise it
must cancel, so a change in device state propagates into the quantum-classical resource bill.
\emph{The decision variables span quantum and classical resources, and the objective is
non-stationary}---which is why one characterization does not suffice, and why profiling must be
continuous beyond a one-time calibration.

Unfortunately, many existing frameworks (e.g.,~\cite{qos-paper,qvm-paper}) 
    optimize the QPU, not the pipeline.
Pipeline-aware frameworks~\cite{xacc,qfw-paper,ibm-qrmi,mqss,pilot-quantum,scim-milq,qumod,qurator,fluence}
focus on orchestrating quantum-classical execution through scheduling, backend management, and
job execution as part of the effort to integrate quantum devices with classical HPC
infrastructure~\cite{ref-arch-qcsc,ornl-qcsc-vision,ornl-amir-vision}. \emph{They provide
relatively little visibility into the interactions between individual pipeline stages, classical
resource behavior, and quantum hardware state.} As a result, decisions such as
qubit mapping, QEM selection, and resource provisioning are typically made
independently or statically, despite being tightly coupled through both execution fidelity and
resource consumption. Table~\ref{tab:related} in Section~\ref{sec:related-work} places existing
systems along the axes this coupling implies. 

Our empirical characterization, using QEM as the representative stage, makes the implications
concrete. QEM techniques differ by orders of magnitude in classical compute and memory
demand; the classical phases within a single mitigation run differ in kind (memory-bound vs.
compute-bound) and no single average utilization describes them; the classical host can
sit idle in polling loops for \emph{up to 92\% of wall-time} while waiting on the device, then
spend $25\text{--}290\times$ longer in classical mitigation than the quantum execution itself;
switching QEM techniques swings the fidelity of the same circuit by more than 50\%; and
qubit layout alone can collapse one technique's fidelity to near zero while inflating another's
QPU time by $1.8\times$. Few of these behaviors are visible in the per-circuit measurements that current
frameworks report, and few of the underlying choices can be fixed correctly ahead of execution.
Section~\ref{sec:motive} elaborates these observations in detail.

In this paper, we present LLQM (Low-Level Quantum Machine), a profiling-driven
meta-framework for quantum-classical pipelines. Rather than introducing another workflow
runtime, \brand provides a common profiling substrate that continuously characterizes pipeline
stages together with the hardware state under which they execute. It decomposes each application
into fine-grained typed tasks along a five-stage pipeline: 
    classification,
    pre-processing,
    job queuing,
    execution, 
    and post-processing. 
LLQM records each task's classical and quantum resource behavior against the calibration snapshot it uses 
    as a reference, and feeds telemetry back through two loops: 
    (1) an intra-job loop that revisits a decision when fidelity falls short, 
    and (2) an inter-job loop that keeps the cost models updated with the latest device states 
    for subsequent jobs.
Since tasks are independent by construction, a resource profile can be attributed 
    to one stage of one technique on specific circuits, rather than averaged across a job---which 
    is what makes the task graph a measurement abstraction.

Atop the profiling substrate, \brand defines common interfaces for decisions that consume 
    real resources: hardware-aware qubit mapping via an Expected Success Probability (ESP) metric, 
    QEM technique and configuration selection, 
    and classical resource provisioning, taking
    fidelity targets and resource budgets as inputs. 
\emph{\brand does not prescribe the policy behind any decision.} It provides the runtime
abstraction under which hardware-, fidelity-, and workload-aware policies can be implemented,
evaluated, and compared on identical workloads, hardware states, and budgets.
We discuss the LLQM architecture in Sections~\ref{sec:overview} and~\ref{sec:design}, 
    and characterize the profiling substrate on QASMBench
    circuits~\cite{qasm-bench} of up to $100$ qubits and $10^{7}$ transpiled gates on
    IBM Heron~r2 processors, using mirror circuits~\cite{mirror-circuits} to evaluate fidelity at
    circuit sizes beyond the reach of state-vector simulation.

The paper makes the following contributions:
\begin{enumerate}
    \item We identify \emph{continuous profiling as a missing runtime primitive} for
    quantum-classical pipelines, and show why static optimizations are insufficient
    when stage behavior depends on dynamic device state;
    \item We develop \brand, a profiling-driven meta-framework that
    decomposes a pipeline into typed tasks and records per-phase classical and quantum resource
    behavior alongside runtime device states, with common interfaces for hardware-,
    fidelity-, and workload-aware policies;
    \item Using LLQM, we present a characterization of QEM, 
        exposing the interactions among hardware state, mitigation effectiveness,
        qubit layout, execution fidelity, and quantum-classical resource consumption.
    \item LLQM is maintained as an open-source project.
\end{enumerate}

\section{Motivation}
\label{sec:motive}

We present three case studies to motivate continuous resource
profiling and holistic optimization for
quantum-classical pipelines. 
Each study was conducted with \brand
across circuits ranging from $4$ to $98$ qubits and $\sim$$400$ to $\sim$$10^7$ transpiled gates over
real IBM QPUs.
The profiling of individual tasks across QEM configurations, hardware
states, and qubit mappings made these measurements possible. Together
they show that seemingly local configuration choices---which QEM
algorithm to run, what parameters to choose, and which
physical qubits to use---have first-order effects on both fidelity and
quantum-classical resource usage, and that none of these choices can be made well without
dynamic profiling. 

\begin{figure}[t]
    \centering
    \begin{subfigure}[b]{0.49\columnwidth}
        \centering
        \includegraphics[width=\linewidth]{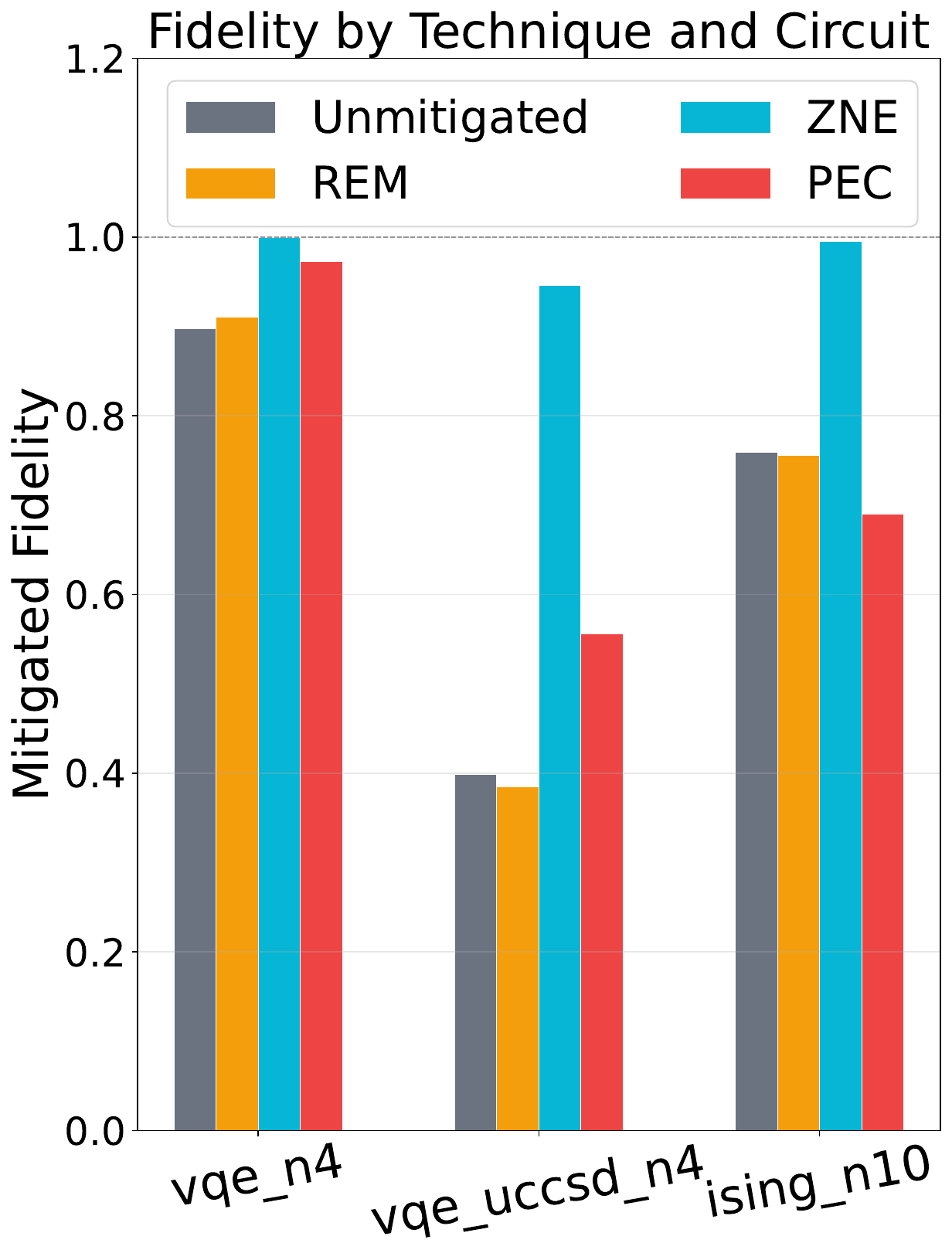}
        \caption{Fidelity of different techniques (ZNE/PEC/REM)
        on Ising and VQE circuits of different sizes.}
        \label{fig:fidelity_comparison}
    \end{subfigure}
    \hfill
    \begin{subfigure}[b]{0.49\columnwidth}
        \centering
        \includegraphics[width=\linewidth]{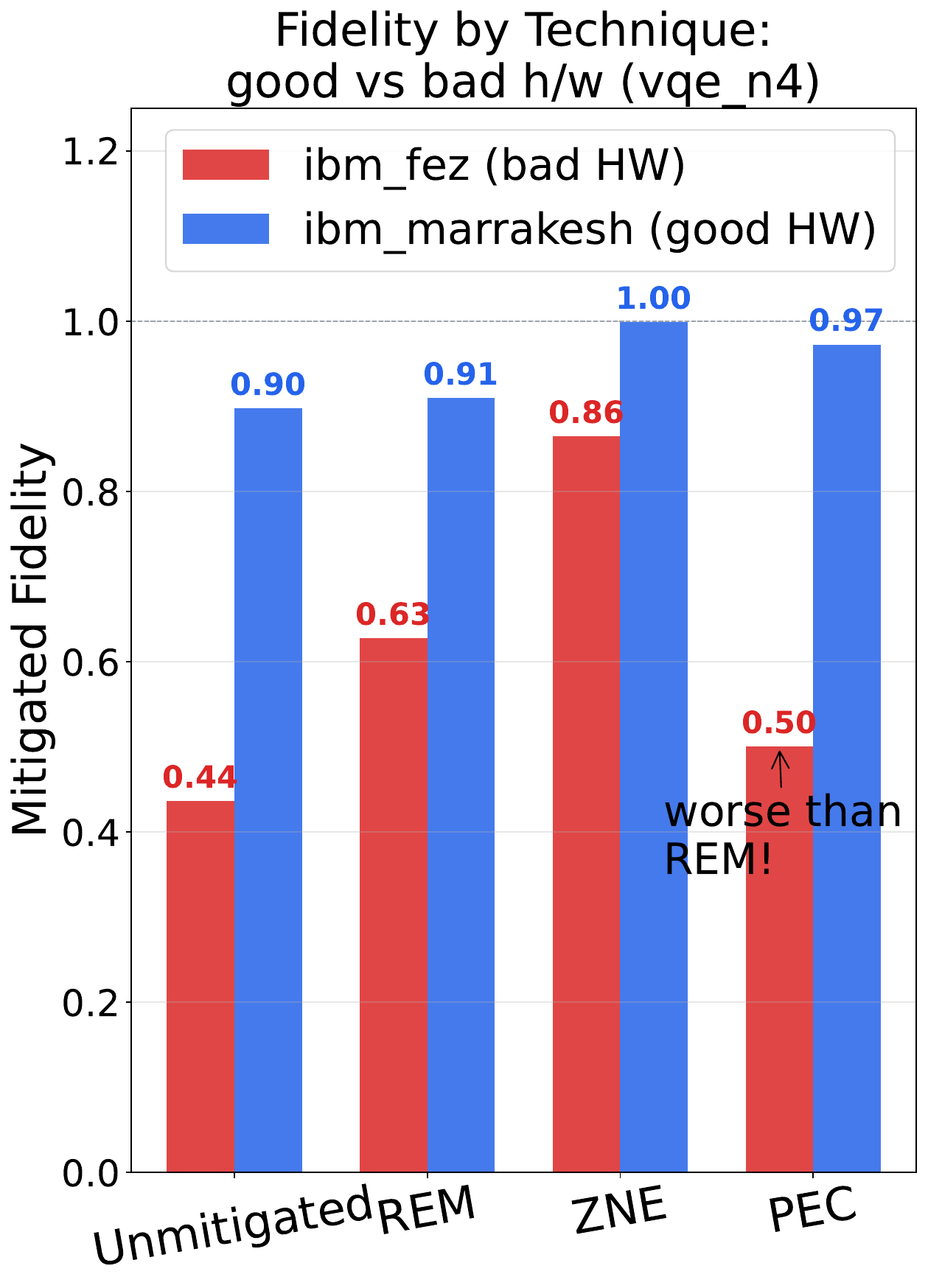}
        \caption{Fidelity of different techniques for the \texttt{vqe\_n4} circuit
        in different hardware states.}
        \label{fig:vqe_fidelity_comparison}
    \end{subfigure}
    \caption{QEM algorithm choice causes large workload- and noise-dependent
    fidelity swings: (a) A QEM algorithm's fidelity varies widely with the type and
    size of the workload and may even fall below the unmitigated baseline; 
    (b) The fidelity of the QEM algorithms
    changes with a slightly different hardware state: PEC performs worse
    than REM on the circuit.}
    \label{fig:no-dominant-technique}
\end{figure}

\para{The choice of QEM algorithms drives major,
non-uniform fidelity swings.}
Figure~\ref{fig:no-dominant-technique} shows that QEM results
    are non-uniform, non-deterministic, and workload- and noise-dependent;
    applying QEM does not guarantee fidelity gains, and misconfigured techniques
    can greatly degrade fidelity. 
As shown in Figure~\ref{fig:no-dominant-technique}(a), configured open-source Zero-Noise
Extrapolation (ZNE)~\cite{em-short-depth-ibm}
achieves consistent fidelity gains across VQE and Ising circuits on an IBM~156-qubit Heron~r2 processor,
whereas Probabilistic Error Cancellation (PEC)~\cite{em-short-depth-ibm,pec-nisq} degrades fidelity on
bigger Ising workloads (\texttt{ising\_n10}). Circuit width, gate count, and algorithm family
(e.g., VQE or Transverse-Field Ising Model) play important roles in QEM algorithm choice and configuration.

Figure~\ref{fig:no-dominant-technique}(b) demonstrates
that spatial and temporal variations in QPU hardware state (comparing a good set of hardware qubits
to slightly degraded hardware qubits) drastically change the performance of the techniques.
On slightly degraded hardware qubits, open-source PEC's fidelity collapses to $0.500$---performing
worse than simple readout error mitigation (REM) ($0.628$)~\cite{measurement-errors} on the
\texttt{vqe\_n4} circuit. The gap between the best and worst QEM choices often determines
    the usefulness of the result, yet the optimal choice cannot be predicted statically
from the circuit alone---it depends heavily on the workload and the device's real-time
noise profile.


\begin{figure}[t]
    \centering
    \includegraphics[width=0.86\columnwidth]{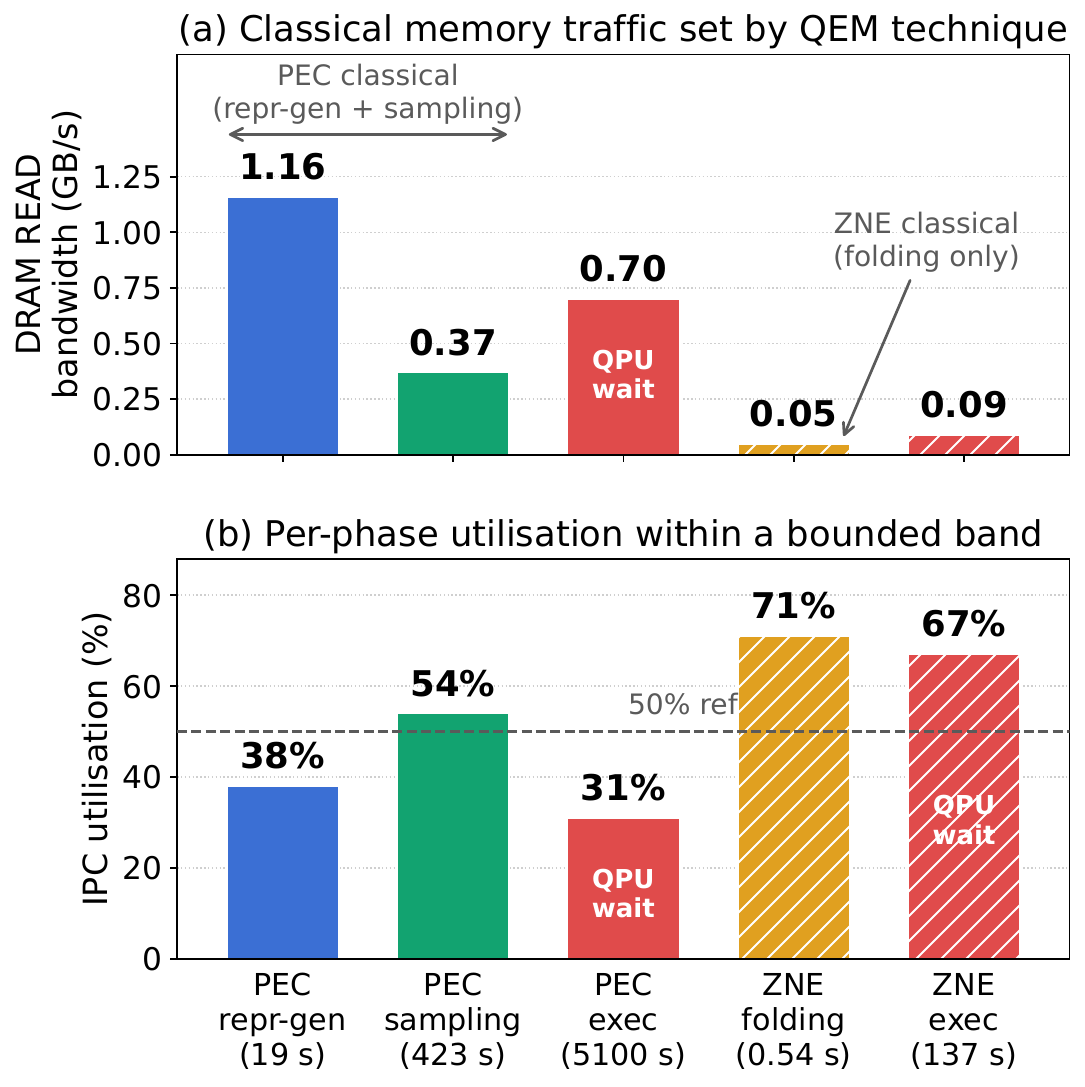}
    \caption{The QEM algorithm determines the classical
    memory/compute footprint: PEC is much more resource-hungry than ZNE, and
    PEC's classical phases move substantial DRAM traffic while ZNE's are negligible.
    Phases differ in compute nature: some are memory-bound (PEC representation generation),
    others compute-bound (PEC sampling); the QPU wait phase polls the QPU for results
    (\texttt{vqe\_n4}).}
    \label{fig:classical_footprint}
\end{figure}

\para{QEM algorithms impose drastically different classical
computing resource costs.}
    Figure~\ref{fig:classical_footprint} shows that error mitigation algorithms
    exhibit vastly different classical compute and memory bottlenecks that dominate
    total execution cost.
While both PEC and ZNE are ``pre-heavy'' techniques, ZNE is a lightweight technique where classical folding
takes less than one second (for \texttt{vqe\_n4}) and the runtime is dominated by QPU execution, whereas PEC requires
a classical sampling overhead $\gamma^2$ that scales exponentially with circuit fault rate:
\begin{equation}
\gamma = \prod_{k \in K} \exp(2\lambda_k) = \exp\left(2 \sum_{k \in K} \lambda_k\right)
\label{eq:pec_gamma}
\end{equation}
where $\lambda_k$ are the learned parameters of the sparse Pauli-Lindblad noise
model~\cite{em-short-depth-ibm,pec-nisq}. However, PEC can fully cancel noise bias, but it can
consume $25\text{--}290\times$ more total execution time than ZNE for the same circuit.
Profiling reveals stark phase-dependent resource characteristics:
PEC representation generation is memory-bound and NUMA-sensitive (moving up to
$1.16$~GB/s DRAM read traffic at $38\%$ physical IPC), circuit sampling is
compute-bound and highly parallelizable ($54\%$ IPC over $423$ seconds), and
the QPU execution/queuing phase stalls classical host CPUs in polling loops for
$5,100$ seconds ($92\%$ of total wall-time) at low IPC ($\approx{}31\%$).
Since per-phase intensity stays within a bounded band,
an algorithm's footprint is predictable from a short profiling window;
however, PEC's aggregate memory grows with the number of sampled circuits,
making memory a first-class, algorithm-dependent budget dimension that a profiler must
account for.
Existing schedulers fail to account for heterogeneous classical burdens,
wasting classical 
resources during QPU queue polling. 


\begin{figure}[t]
    \centering
    \begin{subfigure}[b]{0.49\columnwidth}
        \centering
        \includegraphics[width=\linewidth]{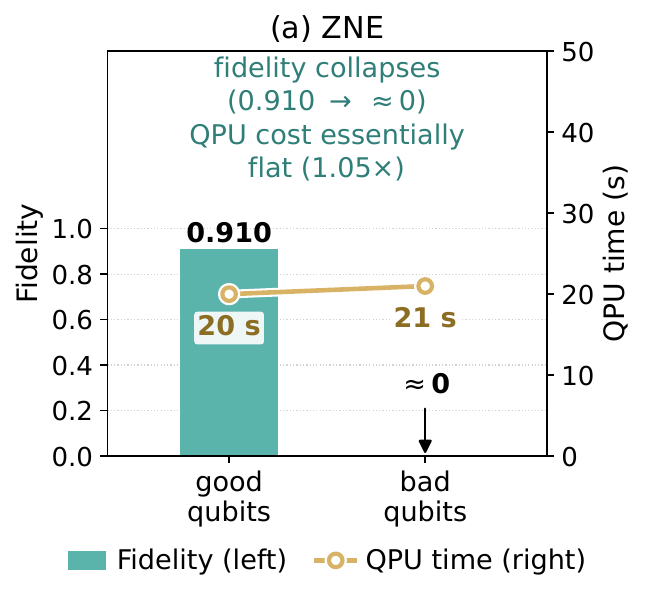}
        \caption{ZNE's fidelity collapses to 0.}
    \end{subfigure}
    \hfill
    \begin{subfigure}[b]{0.49\columnwidth}
        \centering
        \includegraphics[width=\linewidth]{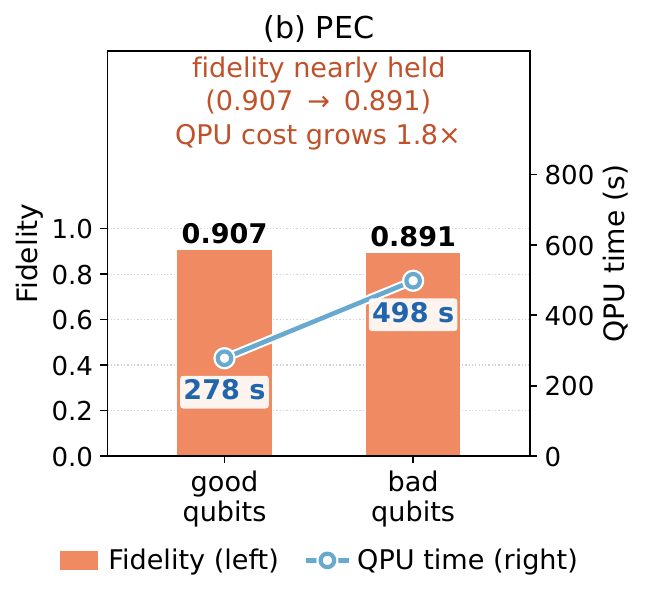}
        \caption{PEC's QPU cost $1.8\times$.}
    \end{subfigure}
    \caption{Qubit selection strongly affects both fidelity and QPU
    execution time: poor qubits collapse ZNE's fidelity and inflate PEC's
    QPU cost (4-qubit circuit); note the change in QPU-time y-axis of the
    two figures.}
    \label{fig:qubit-selection}
\end{figure}

\para{Physical qubit selection and mapping dictate both fidelity and
the scale of quantum-classical costs.}
Figure~\ref{fig:qubit-selection} shows that physical qubit layout has first-order
    impacts on both execution fidelity and QPU time---directly impacting
    quantum-classical resource costs.
Mapping a 4-qubit state-preparation circuit onto
degraded physical qubits causes ZNE fidelity to collapse
from $0.910$ to near zero, rendering the QPU run useless.
Conversely, PEC recovers high fidelity ($0.891$) on the degraded qubits, but does so
at the cost of inflating QPU execution time by $1.8\times$ (from $278$s to $498$s)
due to the higher sampling overhead $\gamma^2$ required to cancel the increased
physical noise (more noise $\rightarrow$ larger $\gamma^2$ $\rightarrow$ more samples
$\rightarrow$ more QPU time). Despite this, the impacts of hardware-aware qubit selection on fidelity
and resource cost remain comparatively under-studied. \brand treats qubit selection as a
first-class optimization target alongside QEM selection and resource
provisioning to prevent cascading resource inflations.
\section{Overview}
\label{sec:overview}

\begin{figure*}[t]
    \centering
    \includegraphics[width=\textwidth]{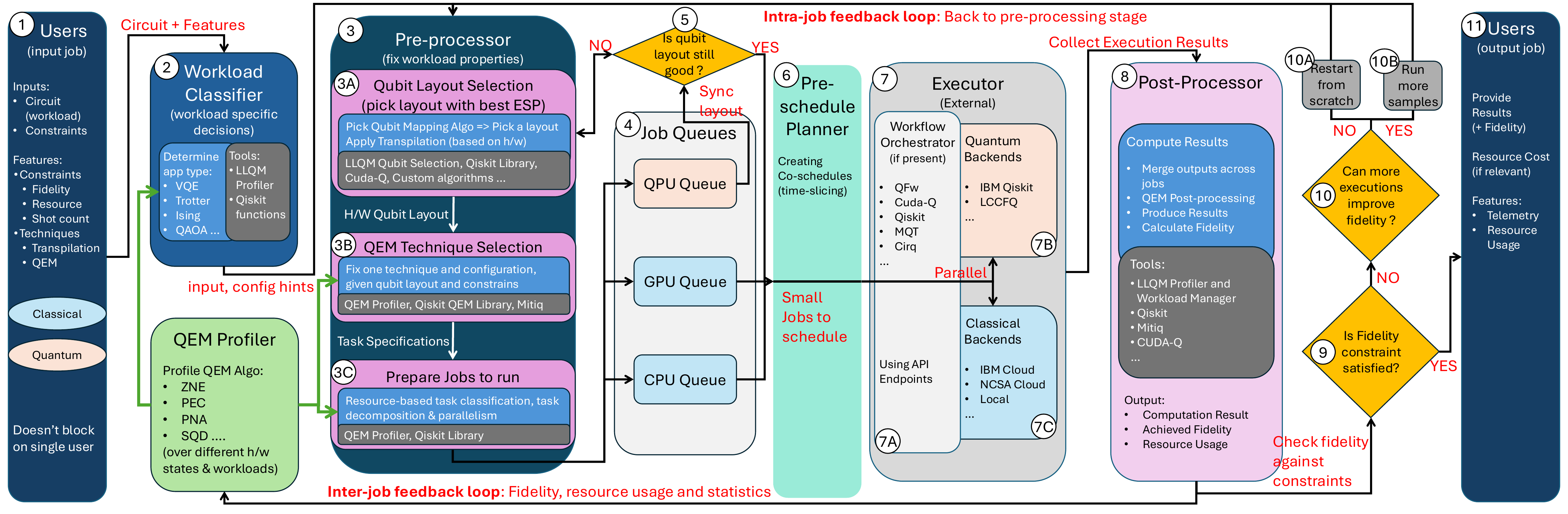}
    \caption{\brand decomposes a quantum-classical pipeline into fine-grained tasks, dynamically profiling each to understand
    local bottlenecks and their implications for dependent operations. The framework holistically improves resource
    efficiency by optimizing configurations (e.g., QEM algorithms), resource provisioning, and hardware-aware
    qubit mapping.}
    \label{fig:overview}
\end{figure*}

\brand is an extensible framework for continuous profiling and
closed-loop optimization of quantum-classical pipelines.
Figure~\ref{fig:overview} shows its components and workflow. 
\brand takes user-specified fidelity targets and resource budgets as inputs,
    and jointly optimizes quantum-classical hardware resources
    by dynamically tuning pipeline stages based on the structure of
    the workload and the noise state of the device at execution time.
    Any stage can plug into \brand through a common interface by declaring
    its configuration space. Our current implementation focuses on qubit layout
    and on QEM technique and configuration---because these decisions dominate both fidelity
    and resource cost on today's devices (Section~\ref{sec:motive}).


\para{Closed-loop continuous profiling.}
\brand establishes a \emph{continuous closed-loop workflow}.
It profiles individual task execution across classical nodes and QPUs and
monitors QPU calibration drift in real time. 
It then dynamically adjusts QEM configurations, layout mappings, and job schedules using 
    the telemetry. 
Two feedback edges close the loop on different timescales. The \emph{intra-job}
edge sends a fidelity verdict back to the decisions that produced it
(steps~\circled{9}$\rightarrow$\circled{10}$\rightarrow$\circled{3}). The \emph{inter-job} edge returns
fidelity, resource usage, and execution statistics from every completed
task to the profiler (step~\circled{8}~$\rightarrow$~QEM profiler), calibrating the cost model that the
\emph{next} job's decisions consume against hardware in a similar noise state.

\para{Constraints and inputs.}
A job enters step~\circled{1} as a quantum circuit $U$ (e.g., OpenQASM~\cite{openqasm} or 
Qiskit~\cite{qiskit}), optionally with hints steering method selection and hard constraints: 
target fidelity $F_{\min}\in[0,1]$ (e.g., $\ge0.80$), quantum time budget $T_q$, 
classical compute budget $T_c$, memory ceiling $\mu_{\max}$, and shot-count caps. 
$T_c$ covers all CPU and GPU time consumed by the job: 
though \brand types every task by the processor it runs on and provisions CPU/GPU separately (Section~\ref{subsec:provision}), 
both count against the same budget to prevent policies from bypassing limits by shifting work between processors. 
These classical budgets may be unnecessary for single-user deployments but are important for the shared systems \brand targets.

\para{Pipeline.}
\brand routes each job through a fine-grained pipeline of five stages,
spanning the eleven steps of Figure~\ref{fig:overview}:
\begin{enumerate}
\item \textit{Workload classifier} \circled{2} admits the circuit, checks user
constraints, identifies the workload family (e.g., VQE, Ising, Trotterized
dynamics, QAOA), and produces a \emph{ranked list} of candidate
QEM and qubit-mapping algorithms. Ranking must happen here because
later stages see only a circuit, and the workload family is what
determines whether a technique applies---SQD~\cite{ibm-riken-chem-sqd},
for instance, exploits chemistry-specific structure that no inspection of the
gate list recovers.
\item \emph{Pre-processor} \circled{3} decides what to run.
It selects a backend and physical qubit layout using an Expected Success
Probability (ESP) metric \circled{3A}, then commits to a QEM technique and
configuration for that layout \circled{3B}. Both decisions are made against a
calibration snapshot, and are only effective when the snapshot is fresh.
The committed configuration
is decomposed into a directed acyclic graph of typed sub-jobs (memory-bound,
GPU/CPU-parallel, or QPU) and pushed to resource queues \circled{3C}.
\item \emph{Job queues and pre-schedule planner} \circled{4}--\circled{6} hold ready
work in heterogeneous CPU, GPU, and QPU queues \circled{4}.
Queued tasks are monitored continuously: if drift invalidates the layout
a task was planned for, that layout and its dependent sub-jobs are
reconfigured \circled{5}. The planner then co-schedules independent
classical and quantum tasks, in parallel or time-sliced across workers, ensuring
neither resource remains idle \circled{6}. Ready tasks are admitted to a
ready set that is global across \emph{all} jobs rather than per-job, so the
scheduler may interleave tasks from different jobs and defer work when
hardware is unavailable.
\item \emph{Executor} \circled{7} dispatches tasks to external quantum backends
(e.g., IBM Quantum and LCCFQ; \circled{7B}) and classical HPC resources
\circled{7C}, directly or via an orchestrator \circled{7A} such as QFw~\cite{qfw-paper} or
CUDA-Q~\cite{nvidia-cudaq}.
\brand exposes a uniform backend abstraction so
every task returns a canonically formatted record regardless of where it ran.
\item \emph{Post-processing and feedback} \circled{8}--\circled{11} fuse job outputs, apply
the selected QEM's estimator, and compute fidelity \circled{8}. If
constraints are met \circled{9}, the result returns to the user \circled{11};
if a shortfall can be
addressed by more sampling~\circled{10}, the job is resubmitted with a larger shot
budget \circled{10B}; otherwise it is restarted \circled{10A},
or fails if no configuration can satisfy the constraints.
\end{enumerate}

\section{Main Components}
\label{sec:design}


\if 
The closed loop of LLQM commits compute resources at
    three points: which physical qubits to run on, which QEM
    technique and configuration to apply, 
    and how much quantum-classical hardware to provision. 
Section~\ref{sec:motive} showed that none of the three decisions
    can be fixed statically, because each
    depends jointly on the structure of the workload and on the noise
    profile of the device at execution time. 
We take them in turn---layout
selection (Section~\ref{subsec:layoutsel}), QEM configuration
(Section~\ref{subsec:qemsel}), and resource provisioning
(Section~\ref{subsec:provision})---preceded by the profiling substrate
they rely on (Section~\ref{subsec:profiling}) and followed by the
mechanism that closes the loop (Section~\ref{subsec:closing}).
LLQM provides extensible \emph{interfaces} and the \emph{measurements}, 
    and leaves the \emph{policy} pluggable: the built-in policies \brand ships derive their
value from making competing policies directly comparable on
workloads, hardware states, and budgets.
\fi

\subsection{Profiling Substrate}
\label{subsec:profiling}

\brand decomposes a job into tasks
    so that it can optimize task executions based on 
    circuit size, noise profile, and resource availability.
Fine-grained task decomposition in \brand is not only an engineering discipline
    but also a precondition for effective optimizations.

\para{Tasks as the unit of measurement.}
\brand compiles a job into a directed acyclic graph of \emph{tasks}. Each task is an atomic operation, 
typed as QPU, CPU, or GPU work, that consumes a fixed set of inputs and produces outputs feeding its child tasks.
Tasks are independent by construction: given its inputs, a task can complete on its own, with no dependency on any other task. 
A single terminal task returns the job's result. Independence is what makes the
graph a measurement abstraction and not merely a scheduling one---a resource
record attributed to a task belongs to one phase of one technique on one
circuit, instead of being averaged away across a job.

\para{From profiles to cost models.}
Table~\ref{tab:profile} lists the information recorded by the profiler.
Section~\ref{sec:motive} shows that per-phase resource intensity remains
within a bounded range, enabling a short profiling window to calibrate the
cost model
$\hat{r}(c \mid P, L, B_t) = (t_q, t_c, \mu)$.
Here, $c$ is a candidate configuration (a QEM technique with parameters and
shot count from Section~\ref{subsec:qemsel}), while $(P,L,B_t)$ denote program
features, qubit layout, and device calibration state from
Table~\ref{tab:profile}. The model predicts quantum time $t_q$, classical
time $t_c$, and peak memory $\mu$ consumed by $c$ under those conditions.

Two qualifications are important. First, $\mu$ is configuration-dependent:
for example, PEC memory grows with sampled circuits, so memory must be
modeled as a first-class budget rather than folded into time. Second, the
executor is partially opaque: cloud tasks expose queue wait, execution time,
and shots consumed, but not provider-side host counters; local backends and
simulators do not have this limitation. Online, these records provide
estimates for the three decisions below. Offline, they reveal actual
resource bottlenecks---e.g., the $92\%$ polling stall in
Section~\ref{sec:motive}, which no selection policy addresses and existing
frameworks fail to expose.

\subsection{Qubit-Layout Selection}
\label{subsec:layoutsel}

Layout selection consumes $B_t$ and the program's structural features. 
It runs \emph{first} because it changes which QEM configurations are feasible (Figure~\ref{fig:qubit-selection}):
the layout fixes the fidelity the run starts from,
which in turn determines the technique required and the quantum--classical cost that technique incurs.

To compare candidates before committing QPU time, \brand scores each
with two static metrics derived from $B_t$. The \emph{circuit-weighted
average} (CWA) error weights each physical qubit's and coupler's error
rate by how often the layout uses it, so a low-quality qubit the
circuit barely touches does not dominate the score, and a medium-quality qubit
on the critical path is not hidden by good but idle neighbors.
The \emph{expected success probability} (ESP) is the probability
that the circuit executes with no error,
\begin{equation}
\text{ESP} = \prod_{q \in \mathcal{Q}_{\text{used}}} (1 - \epsilon_q)^{N_q}
             \times
             \prod_{c \in \mathcal{C}_{\text{used}}} (1 - \epsilon_c)^{N_c},
\label{eq:esp}
\end{equation}
where $\epsilon_q,\epsilon_c$ are single- and two-qubit error rates and
$N_q,N_c$ count how many times the transpiled circuit acts on physical qubit $q$ or coupler $c$; higher is better. The two are complementary:
ESP is dominated by circuit depth and falls off sharply for large
circuits, making it a weak discriminator between layouts of the same
deep circuit, whereas CWA stays comparable across depths but discards
the compounding effect that makes deep circuits fail. Neither is
meaningful in absolute terms; both only rank layouts of the \emph{same}
circuit against one another.

\brand treats layout selection as a pluggable interface over a portfolio of qubit-mapping
algorithms such as Identity, SABRE~\cite{sabre-algo}, GreedyV/E and
R-SMT~\cite{noise-adaptive-paper}, VQA/VQM~\cite{not-all-qubits-equal},
and JIT~\cite{jit-algo}. It generates candidate layouts, scores them, 
and returns the best one that satisfies the user's constraints, 
falling through the portfolio until one is acceptable or the portfolio is 
exhausted---exhaustion being the failure state of Section~\ref{sec:overview}.
We retain the non-noise-aware baselines
deliberately: aggressive noise-aware transpilation concentrates work
onto a small subset of ``good'' qubits and yields only marginal
mean-fidelity gains over a randomized mapping, which itself exhibits
lower run-to-run variability~\cite{revisiting-noise-adaptive}. If that
generalizes, always paying for the most expensive selection pass spends
classical time for no fidelity benefit. Keeping the baselines lets
\brand detect this per circuit rather than assume it away.


\begin{table}[t]
\centering
\caption{Information recorded by the \brand profiler.}
\label{tab:profile}
\small
\begin{tabularx}{\linewidth}{@{}lXl@{}}
\toprule
\textbf{Group} & \textbf{Quantities} & \textbf{Consumer} \\
\midrule
Program $P$ & Circuit width, one- and two-qubit gate counts, workload family & \S\ref{subsec:qemsel} \\
Device $B_t$ & One- and two-qubit gate error rates, readout error rates, calibration age & \S\ref{subsec:layoutsel} \\
Layout $L$ & CWA and ESP derived from $B_t$ over a specific qubit layout & \S\ref{subsec:layoutsel} \\
Task record & Wall-clock CPU/GPU/QPU time, IPC, DRAM read/write bandwidth, shots, queue wait vs.\ execution & \S\ref{subsec:qemsel}, \S\ref{subsec:provision} \\
\bottomrule
\end{tabularx}
\end{table}

\subsection{QEM Selection}
\label{subsec:qemsel}

At this stage, the pipeline holds a committed layout $L$, the
snapshot $B_t$, program $P$ (Table~\ref{tab:profile}), and the
calibrated cost model $\hat{r}$. \brand decides \emph{how} to mitigate
noisy circuit runs.

Let $\mathcal{M}$ be the set of QEM techniques. Each
$m \in \mathcal{M}$ carries a parameter space $\Theta_m$, and every
execution additionally fixes a shot allocation $s$, so a \emph{QEM
configuration} is a triple $c = (m, \theta, s)$ drawn from
$\mathcal{C} = \bigcup_{m \in \mathcal{M}} \{m\} \times \Theta_m \times \mathcal{S}$,
of which exactly one element is committed before any task is
generated. A configuration induces a predicted fidelity
$\hat{F}(c \mid P, L, B_t) \in [0,1]$ and a predicted cost vector
$\hat{r}(c \mid P, L, B_t) = (t_q, t_c, \mu)$, and the user constraints
of Section~\ref{sec:overview} carve out a feasible set
\begin{equation}
\mathcal{C}_{\text{feas}} = \left\{\, c \in \mathcal{C} \;\middle|\;
    \hat{F}(c) \geq F_{\min} \;\wedge\; \hat{r}(c) \leq (T_q, T_c, \mu_{\max}) \,\right\},
\label{eq:qemsel}
\end{equation}
from which the stage returns the cheapest member,
$c^{\star} = \arg\min_{c \in \mathcal{C}_{\text{feas}}} \operatorname{cost}(c)$,
signaling the failure state when $\mathcal{C}_{\text{feas}} =
\emptyset$. Fidelity is a requirement the user states, not a quantity to
maximize: spending more of a shared QPU than $F_{\min}$ demands is a cost
borne by every other job in the system.

\brand currently supports four options: no mitigation, REM, ZNE, and PEC. 
REM corrects only measurement errors; ZNE is cheap and
broadly applicable but leaves a residual bias no amount of sampling
removes; PEC cancels noise bias given a learned noise model, at a
sampling overhead $\gamma^2$ that grows exponentially with the circuit fault
rate (Equation~\ref{eq:pec_gamma}). 
The no-mitigation baseline is a genuine option: Figure~\ref{fig:heatmap} shows mitigated 
configurations that fall below it in fidelity, 
so choosing not to mitigate is a decision and not merely the absence of one. 
A technique joins the portfolio by supplying
three things---its parameter space $\Theta_m$, a cost model $\hat{r}$
calibrated by the profiler, and an estimator that consumes task
outputs---so $\mathcal{M}$ is open and implementation-defined rather
than fixed by the pipeline; we are adding support for more QEM techniques,
    including PNA~\cite{pna},
    SQD~\cite{ibm-riken-chem-sqd}, and QuePP~\cite{quepp}.


Equation~\ref{eq:qemsel} has three terms; profiling makes two of
them cheap: the constraints are given \emph{a priori} and the cost vector is calibrated.
The fidelity predictor
$\hat{F}$ remains a challenge: no sound construction is known.
Its arguments are clear---workload structure and real-time noise
profile jointly determine the outcome, and the profiler collects
both---but the function itself is not.
A recent community assessment reports that
much of the QEM landscape remains empirical and problem-specific, with
few guidelines on when to apply a given technique~\cite{qem-empirical}.
\brand offers extensible interfaces to plug in different QEM techniques
and configurations---the profiler supplies $(P, L, B_t)$ and $\hat{r}$,
    each QEM technique supplies $\Theta_m$, the constraints supply $F_{\min}$
    and the budgets, and the job generator consumes whichever $c^{\star}$ is returned.


\subsection{Resource Provisioning}
\label{subsec:provision}

Once $c^\star$ in Equation~\ref{eq:qemsel} is fixed, \brand expands it into a
task graph and provisions classical resources for each task. Profiling
provides not only scalar cost but also each task's \emph{intensity
class}. Section~\ref{sec:motive} shows that PEC representation generation is
memory-bound, circuit sampling is compute-bound and parallelizable, and
QPU execution performs no useful classical work while pinning large memory.
These distinctions drive provisioning and are invisible without
per-task profiling.

Two decisions are intertwined. 
Per task, \brand selects placement (CPU or GPU)
and core/memory allocation; globally, it selects the next ready task to run.
The ready set spans \emph{all} jobs, allowing cross-job interleaving and
turning the queue-wait stall and post-processing gap in
Section~\ref{sec:motive} into schedulable capacity. \brand only enforces
dependency order; the scheduler optimizes using the task graph, user
constraints, cost predictions $\hat{r}$, live hardware state, and profiler
records.



\subsection{Closing the Loop}
\label{subsec:closing}

\para{Staleness gate.}
Since selection and execution may be separated by arbitrary delays,
\brand monitors the device while a task waits and re-reads the snapshot
before submission. If $B_t$ drifts enough that the committed layout no longer
meets $F_{\min}$, the task is reconfigured and its graph updated, either by
re-running the decisions of Sections~\ref{subsec:layoutsel} and
\ref{subsec:qemsel} or by returning failure as per deployment policy.
The gate cannot catch drift after submission: Section~\ref{sec:evaluation}
shows PEC campaigns running for hours, long enough for the device to drift away from
the Pauli-Lindblad model learned at startup. Preventing this requires
bounding a technique's sampling duration by its calibration-decay timescale,
which the profiler does not yet measure.

\para{Backend abstraction.}
Quantum tasks are dispatched through a unified interface across cloud
providers, local QPUs, and simulators. This enables comparable profiles:
every execution returns a unified result format (result type, raw and
normalized measurements, achieved fidelity, shots, and time consumed), so
task records have consistent meaning across backends.

\para{Feedback.}
When a job's terminal task completes, its results are fused and checked
against the constraints. If the constraints are satisfied, the result returns to the user. 
If the shortfall looks correctable by more shots or a minor parameter change, 
the job is reconfigured and resubmitted. 
Otherwise, it re-enters the decision stages or fails. Each outcome is also a measurement: a
completed job yields a $(c, L, B_t)$ triple annotated with achieved
fidelity and true resource cost---the supervision the missing predictor
$\hat{F}$ of Section~\ref{subsec:qemsel} would require.
\brand does not learn that predictor, but it is built such that the data
needed to learn it accrues as a by-product of running jobs.

\section{Preliminary Results}
\label{sec:evaluation}

The motivating studies in Section~\ref{sec:motive} showed, on
a handful of representative circuits, that the choices of 
    QEM techniques,
    configuration parameters, 
    and qubit selection have first-order
    effects on fidelity and quantum-classical resource consumption. 
This section extends the experiments across diverse circuits
(Table~\ref{tab:benchmarks}). 
Despite being preliminary, the results reinforce our observations at a larger scale: 
the right configuration cannot be
chosen statically, strengthening the motivation 
of \brand's profiling-driven, closed-loop optimizations.

\subsection{Setup}
\label{subsec:setup}

\noindent
{\bf Benchmark.} We evaluate \brand on benchmark programs from OpenQASM-2~\cite{qasm-bench}, 
    spanning its small, medium,
and large categories and covering VQE, UCCSD-VQE, Ising, and
Trotter families with widths of up to $\sim$$100$ qubits
and up to $\sim$$10^7$ transpiled gates. Table~\ref{tab:benchmarks}
lists each circuit with its native and transpiled gate counts; we report
transpiled counts because the transpiled circuit---not the logical
one---is run on the QPU.

\begin{table}[t]
\centering
\caption{Benchmark circuits used in our studies (from
QASMBench~\cite{qasm-bench}). Native counts are the logical
OpenQASM-2 gates and CNOT counts; transpiled gate
count is measured after mapping onto the IBM Heron~r2 processor.}
\label{tab:benchmarks}
\begin{tabular}{lrrr}
\toprule
\textbf{Circuit} & \textbf{Gates} & \textbf{CNOTs} & \textbf{Transpiled gates} \\
\midrule
\texttt{vqe\_n4}          & 89      & 9       & 414 \\
\texttt{vqe\_uccsd\_n4}   & 220     & 88      & 1{,}832 \\
\texttt{ising\_n10}       & 480     & 90      & 2{,}481 \\
\texttt{ising\_n26}       & 280     & 50      & 1{,}982 \\
\texttt{ising\_n34}       & 368     & 66      & 2{,}463 \\
\texttt{ising\_n66}       & 720     & 130     & 5{,}415 \\
\texttt{ising\_n98}       & 1{,}072 & 194     & 7{,}948 \\
\texttt{basis\_trotter\_n4} & 1{,}626 & 582     & 12{,}739 \\
\texttt{vqe\_uccsd\_n6}   & 2{,}282 & 1{,}052 & 21{,}769 \\
\texttt{vqe\_uccsd\_n8}   & 10{,}808& 5{,}488 & 116{,}974 \\
\texttt{vqe\_uccsd\_n28}  & 399{,}482 & 296{,}648 & 9{,}178{,}441 \\
\bottomrule
\end{tabular}
\end{table}

\para{\bf Hardware measurement.} Quantum workloads ran on IBM 156-qubit
Heron~r2 processors; QPU time is the provider-reported billed duration for
which the QPU was locked. All classical mitigation work (folding, representation
generation, sampling, estimation) ran single-threaded on one core of an Intel Xeon
Platinum~8474C, giving a conservative, parallelization-free per-task cost.
This single-host configuration is deliberate: it establishes the per-phase
resource profile of each technique in isolation, before parallelization or multi-node
placement is applied. The profiling substrate and the three decisions it feeds are
independent of deployment scale; what changes with scale is the difficulty of the
provisioning policy in Section~\ref{subsec:provision}, not the measurements it consumes.
Classical time is instrumented per pipeline phase; DRAM bandwidth and IPC
were collected with Linux \texttt{perf}~\cite{perf} in a separate run so that profiling overhead
would not perturb the reported timings.
To evaluate fidelity without incurring
the exponential cost of classical state-vector simulation, every
circuit $U$ is transformed into a mirror circuit by appending
its inverse $U^{\dagger}$, ensuring that the ideal output
is exactly the all-zeros $|0\ldots0\rangle$~\cite{mirror-circuits}.
The observable $M$ is the spatial average of the single-qubit
Pauli-$Z$ expectation values across all $n$ qubits,
\begin{equation}
M = \frac{1}{n}\sum_{i=1}^{n} \langle Z_i \rangle,
\end{equation}
whose ideal value is $\langle M\rangle_{\text{ideal}} = +1$.
Execution fidelity is its linear distance from the
mitigated expectation value $\hat{M}$, 
\begin{equation}
\text{Fidelity} = \max\!\left(0,\; 1 - |\hat{M} - 1|\right).
\label{eq:fidelity}
\end{equation}

\begin{figure}[t]
\centering
\includegraphics[width=\columnwidth]{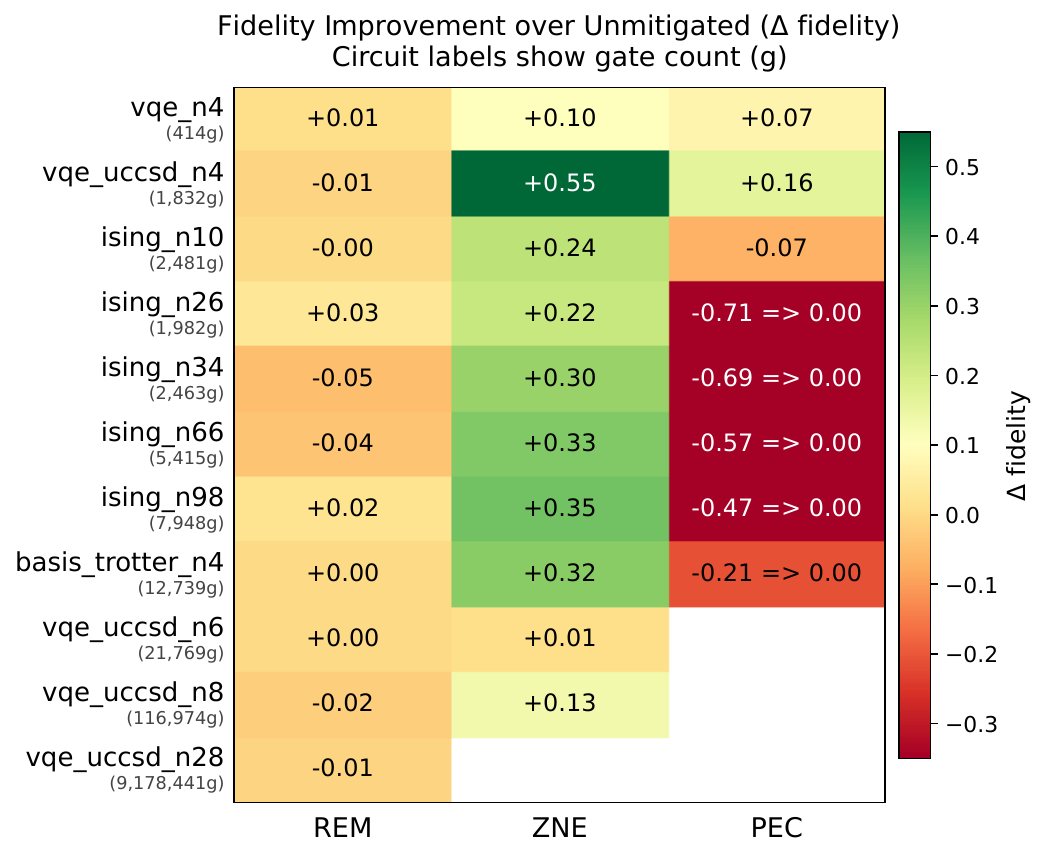}
\caption{Fidelity improvement over the unmitigated baseline
($\Delta$~fidelity) for each technique across
circuits (row label: transpiled gate count $g$). ZNE
delivers consistent gains, REM is near-neutral, and PEC
ranges from improvement on small circuits to severe degradation
as circuit size grows.}
\label{fig:heatmap}
\vspace{-10pt}
\end{figure}


\subsection{Results and Observations} 

\noindent
{\bf No single QEM technique is universally effective;
    an ill-fit technique or suboptimal
    configuration parameters can impair fidelity.}
Figure~\ref{fig:heatmap} reports the fidelity change relative to the
unmitigated baseline ($\Delta$~fidelity) for each technique across
circuits. ZNE configured per circuit improves fidelity consistently
($+0.10$ to $+0.55$). REM is near-neutral and occasionally
negative, as measurement-error mitigation cannot address a
gate-error-dominated regime. PEC improves the two smallest circuits and
then collapses fidelity to $0.00$ on every larger circuit it ran on.
One reading of the figure worth guarding against: $\Delta$ understates
PEC's harm as circuits grow, because the unmitigated baseline itself
falls with gate count, so an identical collapse to $0.00$ registers as a
progressively smaller $\Delta$ ($-0.71$ on \texttt{ising\_n26} down to
$-0.21$ on \texttt{basis\_trotter\_n4}). On the other hand, recall from
Figure~\ref{fig:qubit-selection}, with a slightly
degraded hardware state, ZNE's fidelity collapses, but PEC is able to recover.
For a large fraction of circuits, the wrong technique results 
    in worse fidelity than no mitigation;
    the best choice is not predictable from the circuit family
    alone, showing the importance of profiling-driven QEM selection. 

\begin{figure}[t]
\centering
\includegraphics[width=\columnwidth]{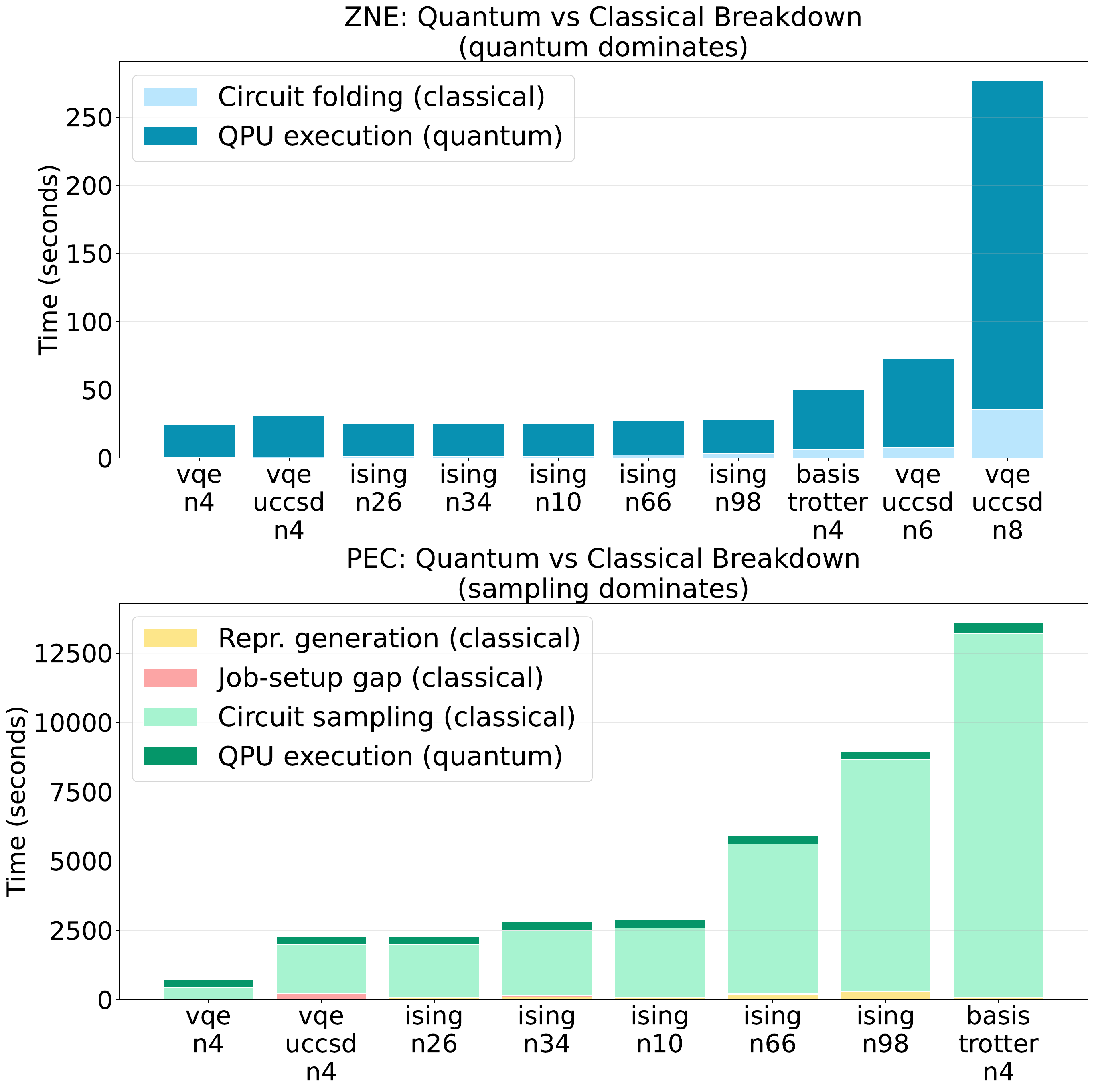}
\caption{Breakdown of execution time of ZNE and PEC into
quantum and classical components, across various circuit sizes.}
\label{fig:breakdown}
\vspace{-7.5pt}
\end{figure}

\para{The classical resources, not the QPU, can be the execution bottleneck;
the cost is QEM-technique-dependent.} Figure~\ref{fig:breakdown} breaks down
the execution of ZNE and PEC into quantum and classical components.
(Note that the \textit{timescales} are different.)
The execution time of ZNE is modest and is dominated by the QPU time;
the execution time of PEC, on the other hand, is overwhelmingly dominated by the CPU time: 
    circuit sampling reaches
    $\sim$$1.3\times10^{4}$~s, which is 
    $25$--$290\times$ of ZNE cost for the same circuit. 
QPU time is irreducible by provisioning; PEC's classical cost is not.
The bottleneck here is \emph{classical} sampling---the host drawing and
constructing the quasi-probability circuits, before any of them reaches the
QPU---and the classical times we report are single-threaded
(Section~\ref{subsec:setup}). They are therefore a per-task cost profile
rather than an achievable latency: the sampled circuits are mutually
independent, so this is precisely the phase a scheduler with spare cores
could collapse. A QPU-centric scheduler does not; it provisions for the
quantum time and leaves this classical work unaccounted for entirely.


\begin{figure}[t]
        \centering
        \includegraphics[width=\columnwidth]{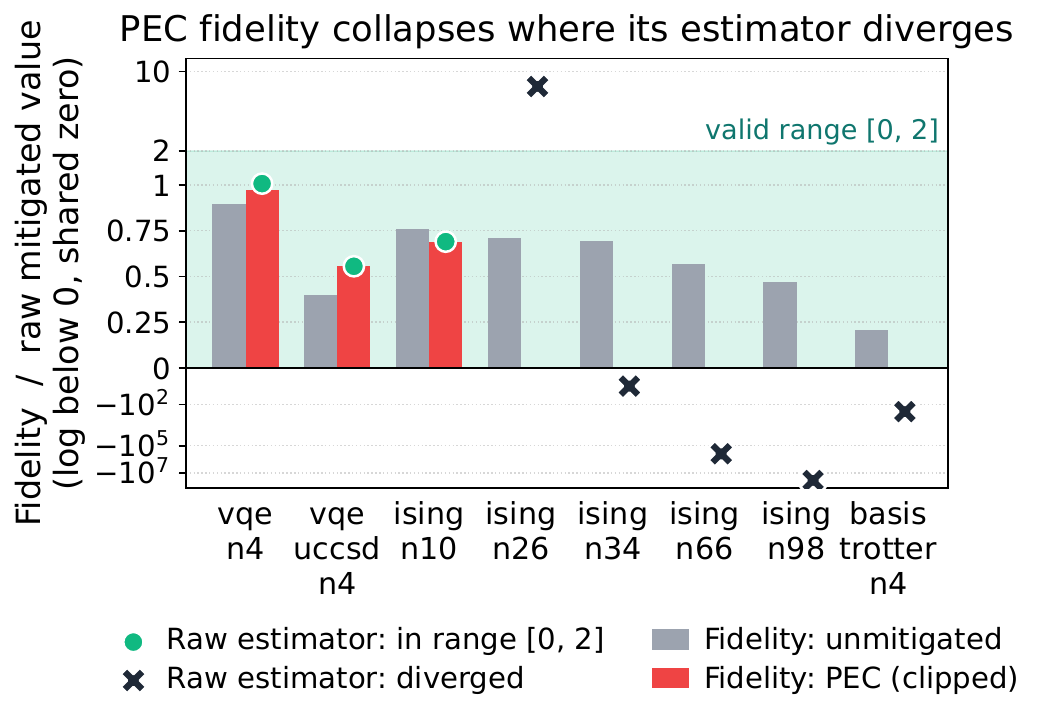}
    \caption{PEC failure mechanism. On large and higher-fault circuits the estimator
    diverges; each PEC point requires a large sampling time ($\sim$3--4 hours), during which hardware drift
    can invalidate the learned noise model.}
    \label{fig:pecdiv}
\end{figure}

\para{PEC's degradation is not random noise but can be explained
    (and avoided).} 
Figure~\ref{fig:pecdiv} explains the PEC collapse in
Figure~\ref{fig:heatmap}. On large circuits with more faults, the raw
(unclipped) estimator diverges far outside the measurement value range $[0,2]$, collapsing fidelity.
The cause is strongly correlated with cost: for these circuits, PEC's sampling time
can be as long as hours,
    long enough for hardware calibration to drift
    and invalidate the learned
    Pauli-Lindblad noise model, so the generated samples no longer cancel the
true noise~\cite{pec-nisq}. PEC's viability is thus tied to both circuit
fault rate and hardware stability, precisely the conditions \brand profiles before
committing to a technique, and monitors for changes.

\begin{figure}[t]
    \centering
    \begin{subfigure}[b]{0.49\columnwidth}
        \centering
        \includegraphics[width=\linewidth]{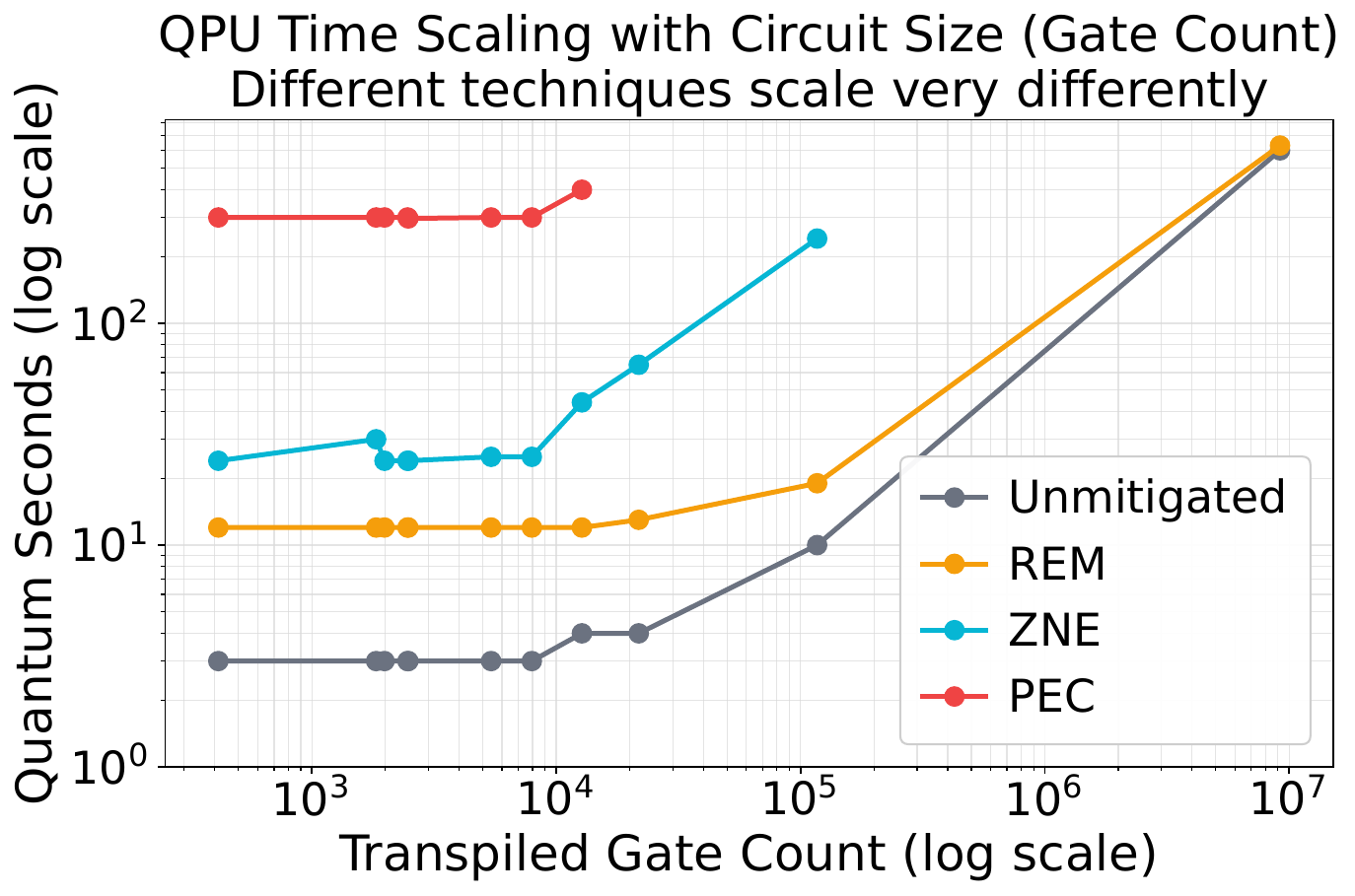}
        \caption{QPU time versus gate count.}
    \end{subfigure}
    \hfill
    \begin{subfigure}[b]{0.49\columnwidth}
        \centering
        \includegraphics[width=\linewidth]{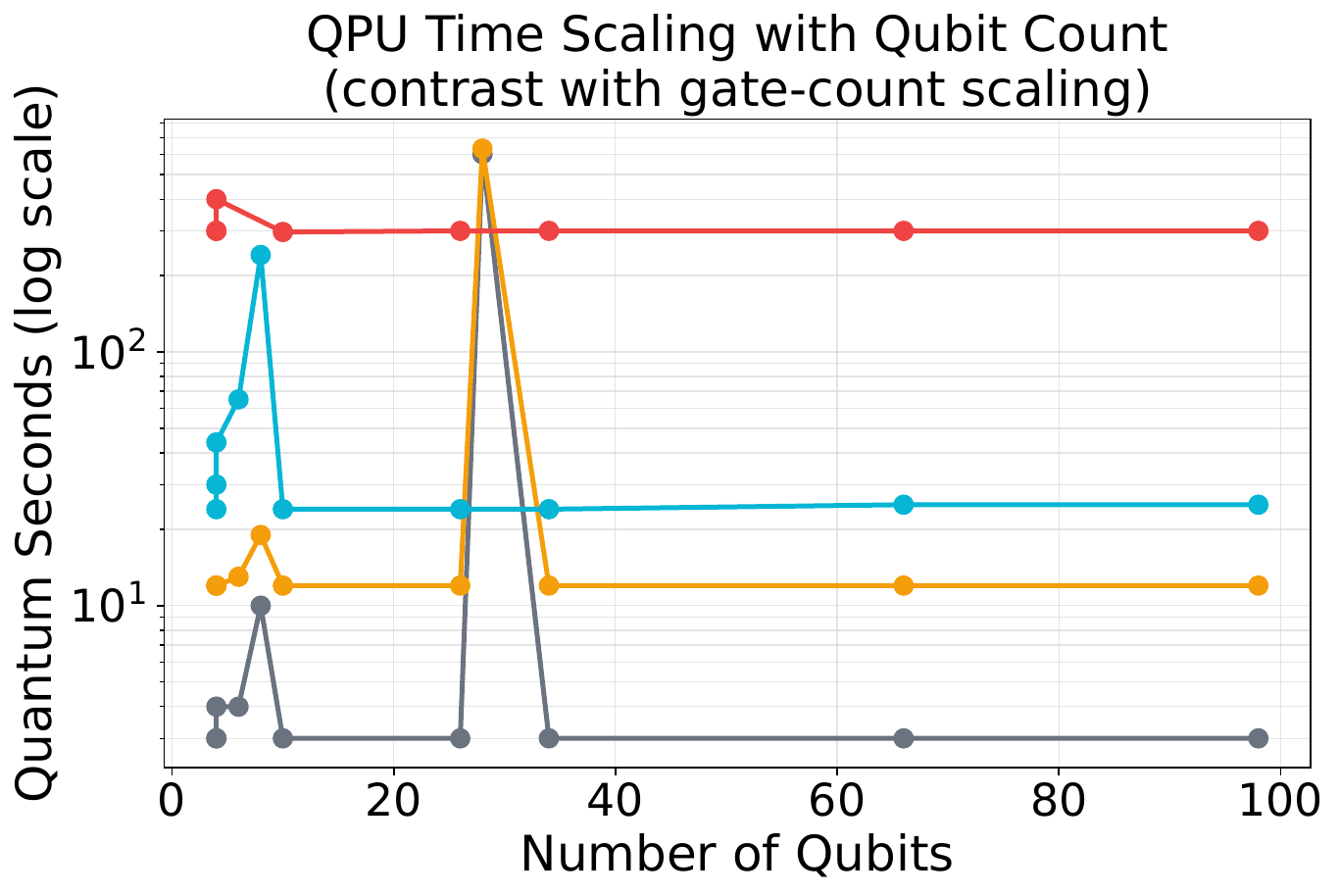}
        \caption{QPU time versus qubit count.}
    \end{subfigure}
    \caption{Transpiled gate count is the primary cost driver for different techniques.}
    \label{fig:timescale}
\end{figure}

\begin{figure}[t]
\centering
\includegraphics[width=\columnwidth]{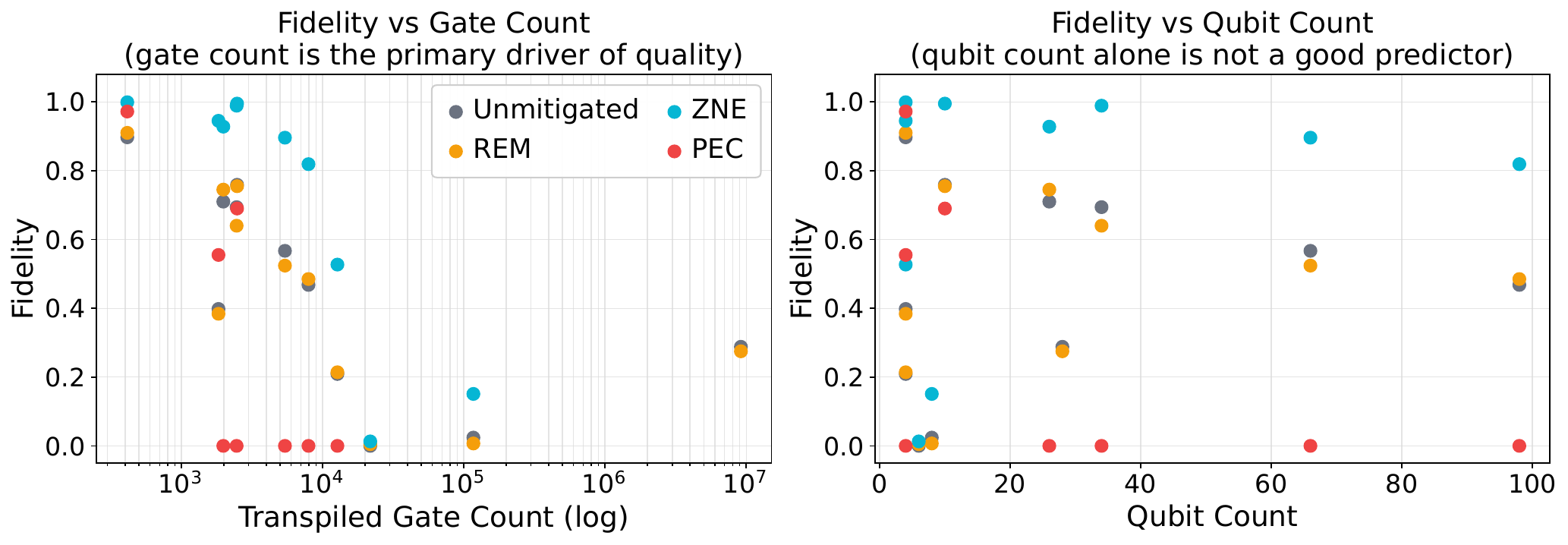}
\caption{Transpiled gate count is the primary fidelity driver for different techniques.}
\label{fig:fidelity-scale}
\end{figure}

\para{Transpiled gate count predicts cost and quality better than qubit
count.} Qubit count is the number a QPU is advertised by, and the one a
scheduler most readily has, but transpilation decouples it from the work
the device actually performs. Measured QPU time follows the transpiled gate
count (Figure~\ref{fig:timescale}(a)): each technique traces
its own near-monotonic trajectory, PEC highest and steepest, so a cost
model keyed on the gate count both separates the techniques from one another
and extrapolates within one. Against qubit count (Figure~\ref{fig:timescale}(b))
the same measurements collapse into a band with no usable trend. Fidelity
behaves the same way (Figure~\ref{fig:fidelity-scale}). The 4-qubit
\texttt{basis\_trotter\_n4} consumes more total execution time than the 98-qubit
\texttt{ising\_n98} (Figure~\ref{fig:breakdown}), and
reaches lower fidelity (Figure~\ref{fig:heatmap}). \brand
therefore uses transpiled gate count as a primary feature for $\hat{r}$
(Section~\ref{sec:design}); sizing the same jobs by qubit count would
mis-provision them by orders of magnitude.


\section{Experience and Discussion}
\label{sec:discussion}

Our experience building \brand exposes several implications for
quantum-classical pipeline optimization. 
We show that fidelity is fundamentally different from conventional resource objectives.
However, the desired outcome cannot be directly observed during production execution,
    and optimization quality improves as the system accumulates workload and device experience. 
In this section, we discuss these implications and 
    clarify the boundary of \brand's current design, 
    including optimizations that fit its abstraction 
    but introduce trade-offs it cannot yet make (which we will 
    discuss in Section~\ref{sec:open}).

\para{\bf Fidelity is a constraint, not a currency.}
Time and memory are tradable---a policy may spend classical cores to save
QPU-seconds, or vice versa. Fidelity is not. Below the target
$F_{\min}$, the result is unusable and its resources are wasted: in
Section~\ref{sec:motive}, a layout choice collapses fidelity to $\approx 0$,
making the QPU time worthless. Above $F_{\min}$, extra fidelity has no value
to a user who set the target, while the QPU time it consumes is expensive.
Thus, when resources are flexible, spending more resources is preferable
to accepting insufficient fidelity.

\para{The computation result is not observable at runtime.}
The fidelity constraint cannot be measured on the production job: its
expectation value is unknown by definition. Mirror circuits provide an
obvious proxy: appending $U^{\dagger}$ produces a circuit with a known
noise-free output (Section~\ref{sec:evaluation}), enabling
fidelity-constrained optimization across configurations. However, the probe
is a different, roughly twice-as-deep circuit with a different observable,
and QEM performance is observable-dependent. It also obscures
workload-specific techniques: $UU^{\dagger}$ destroys the chemistry
structure exploited by methods such as SQD~\cite{ibm-riken-chem-sqd}.
Moreover, each probe incurs full QPU cost, leaving its return on investment
an open question.

\para{Optimizations improve as the profiler collects data.}
Every completed job records its configuration, device state, achieved fidelity,
and cost. Thus, decision-making across pipeline stages gains supporting data
as workloads accumulate, while the same records enable fair comparison of
competing policies under identical workloads, hardware states, and budgets,
rather than the anecdotal comparisons common in prior work~\cite{qem-empirical}.

\section{Open Problems}
\label{sec:open}

LLQM provides the substrate for continuously profiling 
    quantum-classical pipelines to inform three decisions: 
    layout selection, QEM technique and configuration, and quantum-classical resource provisioning. 
However, making optimal decisions remains challenging.
In this section, we discuss these challenges, which are 
    fundamental to any system making these decisions, 
    and are essential for efficient pipeline execution.

\para{Qubit layout selection.}
First, no mapping algorithm is known to be best a priori. Many exist, 
    and \brand maintains a portfolio
    (Section~\ref{subsec:layoutsel}), but its static scores rank their output
layouts, not the algorithms. These scores serve only as
an inexpensive filter for clearly poor candidates. Which algorithm is best
for a circuit remains unknown until its layout is evaluated on the device
state at execution. Even that measurement may not transfer to the next job
if the device has since drifted.

Second, drift can outpace reconfiguration. Before submission, \brand
re-scores a queued task against a fresh snapshot and reconfigures it if its
committed layout no longer meets $F_{\min}$
(Section~\ref{subsec:closing}). If reconfiguration takes longer than the
snapshot's validity, the task can livelock, repeatedly consuming classical
budget without reaching the executor. Avoiding livelock requires choosing
among three costly outcomes: submitting with a stale layout, accepting a
lower-ranked one, or failing the job. Unlike time or memory, fidelity cannot
be traded away (Section~\ref{sec:discussion}). Detecting this condition is
itself an open problem: the profiler does not yet measure the calibration-decay
timescales needed to distinguish persistent rapid drift from an unlucky
sequence of reconfigurations.

\para{QEM selection.}
First, shots and technique cannot be optimized independently. 
The rank order of techniques changes with the shot budget~\cite{united-qem}; 
ZNE crosses a finite-shot boundary below which it degrades the estimate~\cite{zne-help-harm}.
Equation~\ref{eq:qemsel} therefore ranges over whole configurations $c = (m,\theta,s)$ 
    rather than techniques alone, but doing so enlarges the search space.
Hence, $\hat{F}$ must be accurate across shot budgets, not merely across techniques; 
how to calibrate it over that larger space at a tractable profiling cost is an open question.

Second, resources may not have a common price. 
Reducing $\operatorname{cost}(c)$ to a scalar requires an exchange rate among QPU-seconds, 
    CPU-seconds, and memory, and that rate decides 
    whether PEC is selected. 
No amount of profiling supplies the number---it is a valuation, not a measurement---so 
it has to come from deployment policy. 
What that policy should be on a shared system whose users value the two resources differently, 
and whether one rate can serve them at all, remains open.

Third, composition is excluded. The configuration $c^{\star}$ returned by Section~\ref{subsec:qemsel} 
    commits to a single technique, whereas deployed mitigation is often layered. 
    Layering breaks the machinery the stage depends on: the profiler records cost per technique in isolation, 
    and neither the cost $\hat{r}$ of a stack nor its fidelity is known to 
    follow from those of its parts, so a calibrated model does not extend to the composed space, 
    which is combinatorially larger than the single-technique space $\mathcal{C}$. 
How to predict a stack from its components, 
    or profile stacks directly at feasible cost, is unresolved.

Selection is therefore a fundamental runtime responsibility, not merely a
response to immature hardware: sampling overhead for a broad class of QEM
protocols grows exponentially with circuit depth~\cite{qem-sampling-bound},
so no engineering effort removes the need to choose what to run under a fixed budget.

\para{Resource provisioning.}
First, executors differ in observability. As noted in Section~\ref{subsec:profiling},
host-side counters (e.g., queue delays) 
    can be inaccessible outside cloud providers. 
The partial observability limits provisioning decisions, and how to provision
effectively with incomplete executor state remains an open problem.

Second, the profiler starts cold. Cost models rely on task records that
accumulate over time (Section~\ref{subsec:closing}), yet the short profiling
window in Section~\ref{subsec:profiling} presumes sufficient history. New
deployments, backends, or portfolio techniques therefore keep the gap open.
How to make reliable decisions before sufficient history exists, and how
to bound the cost of those early decisions, remain open.
\section{Related Work}
\label{sec:related-work}

\para{Resource orchestration.} 
Prior work \cite{xacc,qfw-paper,ibm-qrmi,mqss,pilot-quantum} 
    orchestrates heterogeneous devices, 
    treating QPUs as manageable resources alongside CPUs and GPUs. 
For example, QFw provides HPC-aware orchestration and distributed simulation \cite{qfw-paper};
QRMI makes QPUs schedulable alongside CPUs and GPUs in workload managers \cite{ibm-qrmi}.
These frameworks execute a specific configuration whose QEM technique, 
    qubit layout, and shot budget are already chosen. 
\brand enables dynamic optimizations by profiling the primitives these frameworks build on, 
    while using them as the execution layer.

\para{Multi-programming.} Prior work~\cite{qos-paper,qvm-paper} improves QPU utilization by sharing a single QPU 
    across multiple programs. 
QOS \cite{qos-paper} jointly optimizes circuits, error mitigation, multi-programming, and spatio-temporal multiplexing, 
    while HyperQ \cite{qvm-paper} provides virtual-machine isolation for multiplexed programs. 
Both apply mitigation uniformly rather than selecting it under a budget, and neither accounts for the classical 
    resources that mitigation consumes. \brand instead makes this cost explicit; its stage interface can compose with multiplexing, 
    but multiplexing trades fidelity for utilization, which \brand's constraint semantics disallow (Section~\ref{sec:discussion}).

\para{Scheduling.} Existing schedulers optimize when and where jobs run to keep quantum and classical resources utilized. 
For example, SCIM MILQ minimizes makespan while reducing noise \cite{scim-milq}, 
QuMod does so across modular QPUs \cite{qumod}, Qurator ranks quantum DAGs using calibration-derived success scores \cite{qurator}, 
and Fluence addresses the two-queue problem \cite{fluence}. Other systems expose QPUs to conventional schedulers 
    without scoring work \cite{slurm-het-quantum,conqure,dqc-scheduling,efaas}. 
Where fidelity enters, it is a static pre-submission score, and none accounts for the classical cost of the chosen mitigation. 
\brand complements rather than replaces these schedulers: it decides what a job is before queuing, 
    while the scheduler decides when it runs using measured quantum and classical costs.

\begin{table}[t]
\centering
\caption{Comparing LLQM with other quantum frameworks.
\pt{}: partial support, e.g., overly coarse
granularity, static compute, or no feedback loop.
{\small Notation: accounting unit (G),
    continuous phase-level profiling (P),
    hardware-aware qubit selection (L),
    QEM selection (M), adaptation to dynamic hardware (D),
    fidelity-aware optimization (F),
    and joint quantum-classical optimization (J).}}
\label{tab:related}
\setlength{\tabcolsep}{6pt}
\begin{tabular}{@{}llccccc c@{}}
\toprule
\textbf{Approach} & \textbf{G} & \textbf{P} & \textbf{L} & \textbf{M} & \textbf{D} & \textbf{F} & \textbf{J} \\
\midrule
\multicolumn{8}{@{}l}{\textit{Resource orchestration}}\\
QFw \cite{qfw-paper}, XACC \cite{xacc}                  & Job      & \no  & \no  & \no       & \no  & \no  & \pt \\
Pilot-Quantum \cite{pilot-quantum}         & Task     & \pt  & \no  & \no       & \no  & \no  & \pt \\
QRMI \cite{ibm-qrmi}, MQSS \cite{mqss}                  & Job      & \no  & \pt  & \no       & \pt  & \no  & \pt \\
\midrule
\multicolumn{8}{@{}l}{\textit{Multi-programming}}\\
QOS \cite{qos-paper}                       & Circuit  & \no  & \yes & Fixed     & \pt  & \yes & \no \\
HyperQ \cite{qvm-paper}, \cite{qvm-tum}    & Circuit  & \no  & \pt  & \no       & \no  & \pt  & \no \\
\midrule
\multicolumn{8}{@{}l}{\textit{Scheduling}}\\
SCIM MILQ \cite{scim-milq,qumod}           & Circuit  & \no  & \pt  & \no       & \no  & \pt  & \no \\
Qurator \cite{qurator}                     & Task     & \no  & \pt  & \no       & \pt  & \yes & \pt \\
Fluence \cite{fluence}, \cite{qpu-sharing-strategies} & Job   & \pt  & \no  & \no       & \no  & \no  & \pt \\
Co-exec \cite{slurm-het-quantum,conqure,dqc-scheduling,efaas} & Job & \no & \no & \no & \no & \no & \pt \\
\midrule
\multicolumn{8}{@{}l}{\textit{Characterization}}\\
Mini-apps \cite{quantum-mini-apps}         & Task     & \pt  & \no  & \no       & \no  & \no  & \pt \\
Perf.\ model \cite{nvidia-hybrid-performance} & Workflow & \pt & \no & \no      & \no  & \no  & \pt \\
QFC \cite{qfc-metric}, \cite{revisiting-noise-adaptive} & Circuit  & \no  & \no  & \no       & \no  & \pt  & \no \\
\midrule
\multicolumn{8}{@{}l}{\textit{Configuration}}\\
Noise-aware \cite{noise-adaptive-paper,not-all-qubits-equal,jit-algo} & Circuit & \no & \yes & \no & \pt & \pt & \no \\
QuanTEM \cite{quantem}                     & Circuit  & \no  & \no  & Selected  & \no  & \pt  & \no \\
ML-QEM \cite{ml-qem}, \cite{gsc-qemit}             & Circuit  & \no  & \no  & Selected  & \yes & \yes & \no \\
\midrule
\textbf{LLQM}                              & \textbf{Task} & \yes & \yes & \textbf{Selected} & \yes & \yes & \yes \\
\bottomrule
\end{tabular}
\end{table}

\para{Characterization.} Prior work measures quantum-classical execution costs under
    specific configurations.
Quantum mini-apps \cite{quantum-mini-apps} benchmark quantum-HPC tasks; 
    a workflow-level performance model \cite{nvidia-hybrid-performance} predicts
    hybrid runtime.
Recent work proposes to measure fidelity per cost \cite{qfc-metric} and shows that cost-aware backend ranking 
    differs from fidelity-only ranking across 12 cloud QPUs. 
A controlled study \cite{revisiting-noise-adaptive} finds noise-adaptive transpilation barely outperforms randomized mapping. 
The last two studies are offline, per-circuit comparisons. 
\brand builds on this work with per-phase, continuous characterization that feeds directly into configuration decisions.

\para{Configuration.} Existing work decides configurations of components independently.
Qubit-selection methods use reliability, variation, or just-in-time re-characterization \cite{noise-adaptive-paper,not-all-qubits-equal,jit-algo},
    while QEM methods select codes, learned surrogates, or online policies under drift \cite{quantem,ml-qem,gsc-qemit}.
Each is a policy, and none couples its choice to the layout it runs on or to the classical budget it spends.
\brand does not replace any configuration technique;
    instead, it defines the interface through which such a policy plugs in,
    jointly accounts for layout and classical budget,
    and records per-job outcomes to compare policies under identical workloads, hardware states, and budgets.

\para{Summary.}
Table~\ref{tab:related} positions \brand in the related work. 
Overall, no existing framework pairs a 
    sub-circuit accounting unit with continuous profiling, and none feeds
    that measurement back into the technique, layout, and shot-budget
    decisions jointly, under quantum \emph{and} classical budgets, against
    a device that drifts between decision and execution.



\section{Conclusion and Future Work}
\label{sec:conclusion}


Treating a quantum job as a black box obscures where fidelity and 
quantum-classical resources are actually lost. Looking inside reveals 
a deeper systems challenge: these pipelines are tightly \emph{coupled}, 
so a local decision can reshape costs and outcomes globally; 
each stage is internally \emph{heterogeneous}, with phases that demand fundamentally 
different resources; and the underlying hardware \emph{drifts}, 
turning every optimization decision into a non-stationary one. 
LLQM addresses these challenges by decomposing jobs into typed tasks and continuously 
profiling them against the calibration state under which they execute. 
In doing so, it shifts quantum pipeline optimization from treating the workflow 
as a fixed recipe to a measurable, evolving system. More broadly, 
it suggests that as quantum applications become more sophisticated, 
the boundary between quantum algorithm design and systems optimization will become increasingly difficult to draw.

Future work follows Section~\ref{sec:open} toward a broader vision of autonomous quantum systems. 
Decision-making should move out of the user's hands: 
techniques, configurations, and layouts should be inferred from the circuit and its constraints, 
so that effective quantum computing does not depend on knowing the intricacies of every mitigation technique or device. 
This intelligence should be separated from any particular backend. 
As quantum technologies and provider stacks diverge, we need abstractions 
that allow optimization policies to be learned once and executed across heterogeneous machines rather than rebuilt for each backend. 
Finally, the system must evolve with the hardware itself. 
Continuous profiling is only a first step toward operating in a world where device characteristics, 
software stacks, and workloads all change. 
Fault-tolerant quantum computing will make this challenge even more pronounced: 
    decoding will introduce latency-sensitive classical computation into the quantum loop, 
    with resource demands and real-time deadlines of its own. 
The long-term question, then, 
is not simply how to optimize a quantum circuit, but how to build systems that can continuously understand, adapt, 
and optimize an evolving quantum-classical computation without requiring humans to anticipate every interaction.

\section*{Acknowledgement}
\label{sec:acknowledgements}

We thank the IBM Quantum Open Plan for QPU time and Richard Welp for classical
infrastructure access. We acknowledge the support of the National Center for
Supercomputing Applications (NCSA) and the IBM-Illinois Discovery Accelerator
Institute (IIDAI). This work was funded in part by NSF CNS-2145295 and OAC-2323116.


\bibliographystyle{IEEEtran}
\bibliography{
    refs/quantum-general,
    refs/quantum-error-mitigation,
    refs/quantum-classical-systems,
    refs/quantum-use-cases,
    refs/quantum-algorithm-specific,
    refs/misc
}

\end{document}